\documentclass[12pt]{article}
\usepackage{latexsym}
\usepackage{hyperref}

\def\cmm#1{}              

\def\tr{{\rm tr}\,}
\def\Tr{{\rm Tr}\,}

\def\half{{1\over 2}}

\def\Det{{\rm Det}}

\thicklines                         
\let\I\i

\def\secteq#1{ \setcounter{equation}{0}
               \renewcommand{\theequation}{#1.\arabic{equation}} }

\def\a{\alpha}

\def\c{\chi}
\def\d{\delta}
\def\e{\epsilon}                
\def\g{\gamma}

\def\i{\iota}

\def\l{\lambda}
\def\m{\mu}
\def\n{\nu}

\def\r{\rho}                    

\def\x{\xi}

\def\D{\Delta}

\def\X{\Xi}

\def\cu{{\cal U}}

\def\Det{{\rm Det}}
\def\Tr{{\rm Tr}}
\def\tr{{\rm tr}}
\def\barp{{\overline{p}}}
\def\hatp{{\hat{p}}}

\def\psibar{{\overline{\psi}}}

\def\ie{{\it i.e.}}

\def\Tr{{\rm Tr\,}}

\def\OUT#1{}
\def\fourth{{\scriptstyle \raise.15ex\hbox{${1\over4}$}}}
\def\ghat{\hat\g_5}
\def\Phat{\hat{P}}

\def\eq#1{(\ref{#1})}

\begin{document}
\hyphenation{fer-mio-nic per-tur-ba-tive pa-ra-me-tri-za-tion
pa-ra-me-tri-zed a-nom-al-ous}

\renewcommand{\thefootnote}{$*$}

\begin{center}
\vspace*{10mm}
{\large\bf Observations on staggered fermions at non-zero lattice spacing}
\\[12mm]
Claude Bernard$^a$,\
Maarten Golterman$^b$\footnote{Permanent address: Department of Physics and Astronomy,
San Francisco State University,
San Francisco, CA 94132, USA}\ and\ Yigal Shamir$^c$
\\[8mm]
{\small\it
$^a$Department of Physics, Washington University\\
St. Louis, MO 63130, USA}
\\[5mm]
{\small\it
$^b$Grup de F{\'\I}sica Te{\`o}rica and IFAE
\\ Universitat
Aut{\`o}noma de Barcelona, 08193 Barcelona, Spain}
\\[5mm]
{\small\it $^c$School of Physics and Astronomy\\
Raymond and Beverly Sackler Faculty of Exact Sciences\\
Tel-Aviv University, Ramat~Aviv,~69978~Israel}
\\[10mm]
{ABSTRACT}
\\[2mm]
\end{center}

\begin{quotation}
We show that the use of the fourth-root trick in lattice QCD with staggered fermions
corresponds to a non-local theory at non-zero lattice spacing, but
argue
that the non-local behavior is likely to go away in the continuum
limit.  We give examples of this non-local behavior in the free theory, and
for the case of a fixed topologically non-trivial background gauge field.  In both
special cases, the non-local behavior indeed disappears in the continuum limit.
Our results invalidate a recent claim that at non-zero lattice spacing
an additive mass renormalization is needed because of taste-symmetry breaking.

\end{quotation}

\renewcommand{\thefootnote}{\arabic{footnote}}
\setcounter{footnote}{0}

\newpage

\vspace{5ex}
\noindent {\large\bf 1.~Introduction}
\secteq{1}
\vspace{3ex}

Staggered fermions \cite{Kogut:1974ag}
have long been in use as a method for formulating the quark
sector of lattice QCD.  The main advantages are that they are relatively inexpensive
when it comes to including sea-quark effects in lattice computations, and that they
have an exact chiral symmetry in the limit of vanishing bare quark mass.
The combination of these two advantages makes
it possible to reach rather low quark masses, which
are
essential for any serious
phenomenological applications of lattice QCD.

These benefits come at a price, however.  A theory with
one flavor of staggered fermion on the lattice yields a theory with four quarks
in the continuum limit.  This is a consequence of fermion species doubling, which
is unavoidable in any situation in which an exact chiral symmetry is preserved on
the lattice.  In modern language, these four quarks per flavor of lattice staggered
fermion are referred to as ``tastes."  Only in the continuum limit does the theory
recover a full $SU(4)$ taste symmetry, whereas at any non-zero value of the lattice
spacing this group is broken to a smaller discrete subgroup \cite{mgjs}.

In principle, the four tastes can be given different masses \cite{mgjs}, but this is
not what is done in practice.\footnote{One reason is that breaking the taste degeneracy
requires additional hopping terms in the lattice action, which, for a generic choice, make the fermion
determinant complex.  Also, the existence of a partially-conserved continuous
chiral symmetry depends
on the choice of mass term.}  Instead, each staggered
flavor (up, down, or strange) is given a single mass, leading to four tastes
of degenerate quarks per flavor.
In order to obtain a theory with only one quark per flavor
appearing in sea-quark loops, one reduces the number
of tastes by taking the fourth root%
\footnote{In the isospin limit,
the up--down sector is represented by a square root of a staggered determinant
with the common light quark mass.}
of the degenerate-mass staggered determinant for
each flavor \cite{Marinari:1981qf}.

This formulation of the sea-quark sector of QCD does not necessarily correspond to
a local field theory at non-zero lattice spacing $a$. The potential lack of locality has been
the cause for much concern recently \cite{SCEPTICS} about the application of staggered fermions
to high-precision hadron phenomenology.   At issue is: (1) whether
the theory is local at $a\not=0$, and (2)
whether the theory, if non-local at $a\not=0$,
becomes local in the continuum limit. An alternative way to phrase the second question is to ask
whether the theory is in the correct universality class.

In this paper, we will argue (Sec.~2) that the theory with the fourth root
of the staggered determinant is indeed non-local at non-zero $a$, but that this does not
imply that the answer to the second question is negative.
We connect the
issue of locality to the role of taste and chiral symmetries.
In Sec.~3, we give some simple examples that
show how the correct local continuum theory may indeed be obtained.
In addition, we demonstrate that recent claims about the properties of staggered fermions
at non-zero $a$, in particular about the renormalization of the bare mass
\cite{ah}, are incorrect. A concluding section summarizes our arguments
and results; while the Appendix collects some useful properties of various
Dirac operators in the taste basis.

\vfill\eject
\vspace{5ex}
\noindent {\large\bf 2.~General considerations}
\secteq{2}
\vspace{3ex}

We begin by giving our general argument.
Suppose that the theory with the fourth root \textit{did} correspond
to a local field theory on the lattice at non-zero $a$.  By definition,
this would require that the two theories differ only by a local
functional.  In other words,
\begin{equation}
\label{4throot}
\Det^{1/4}(D_{stag})=\Det(D)\;{\rm exp}(-\fourth\,\d S_{eff})\ ,
\end{equation}
where $D_{stag}$ is
the staggered Dirac operator,
$D$ is a local
lattice Dirac operator that describes one quark field in
the continuum limit, and
$\d S_{eff}$ is a local effective action for the gauge field.\footnote{
Adams \cite{Adams:2004mf} has recently emphasized that
Eq.~(\ref{4throot}) with $\d S_{eff}$ local is indeed the proper definition of locality
of the rooted theory at $a\ne 0$; requiring $\d S_{eff}=0$ would be too strong.}
Saying that $\d S_{eff}$ is local means that it produces
only effects at the scale of the cutoff.  This would imply that,
apart from a renormalization of the gauge coupling constant, the
presence of $\d S_{eff}$ would not affect the behavior at any physical
length scale that is to be held constant as the lattice spacing is taken to zero.

It is rather easy to see that this set of assumptions leads to a conflict with
what we know to be true about the original staggered theory, \ie\ the one without
the fourth root of the determinant.
Taking the fourth power of Eq.~(\ref{4throot}), we have
\begin{equation}
\label{dets}
\Det(D_{stag})=\Det^4(D)\;{\rm exp}(-\d S_{eff})\ .
\end{equation}
Under our assumption $\d S_{eff}$ is local, and it therefore cannot change the
long-distance behavior of any correlation function.  In particular, it cannot have any
effect on the
Goldstone-boson (GB) masses  predicted by the staggered theory defined
by $D_{stag}$, and those predicted by the theory defined by
\begin{equation}
\label{su4}
D_{4t}=D\otimes {\bf 1}\ ,
\end{equation}
where the second factor is a unit $4\times 4$ matrix, to be
interpreted as the identity matrix in taste space.
The operator $D$ describes a lattice theory with one taste;
in a finite volume, the size of the matrix $D$ is in fact four times
smaller than the size of $D_{stag}$.
Clearly, we have that
$\Det(D_{4t})=\Det^4(D)$, and  the lattice theory defined by $D_{4t}$
has a continuous $SU(4)$ taste symmetry.

We can now compare what we know about the  GB spectrum
of the two theories.  In the theory defined by $D_{4t}$, there will be fifteen
GBs, transforming in the adjoint representation of $SU(4)$,
with possibly a common non-vanishing mass if the operator $D$ violates chiral
symmetry and/or is not massless .  Under our assumption
described above, $\d S_{eff}$ does not change this fact: all long-distance
physics would be contained in $D_{4t}$.

The GB spectrum of the staggered theory is completely different,
irrespective of the value of the staggered bare quark mass.   Of course, in the
continuum limit, one recovers fifteen degenerate (pseudo-)GBs, but at non-zero
lattice spacing, they split up into at least four \cite{ls,numev}, and up to seven \cite{mg},
non-degenerate irreducible representations, consistent with the lattice symmetry group of the staggered
theory.   Indeed, at strong coupling
\cite{ks,saclay2},  there is only one exact GB (at zero quark mass), because of the
exact $U(1)_\e$ axial symmetry \cite{ks}.

It thus becomes clear that our assumption on $\d S_{eff}$ cannot be correct.
The effective action $\d S_{eff}$
has to know about the long-distance effects of
taste-symmetry breaking, and cannot be a local functional of
the lattice gauge field.  Of course,
given a local operator $D$, one can always define $\d S_{eff}$ through Eq.~(\ref{dets}) or
Eq.~(\ref{4throot})
(as long as we consider gauge fields on which $D$ has no exact zero modes,
{\it cf.}\ Sec.~3b),
but what we find is that $\d S_{eff}$ cannot be local.   This shows
that the theory defined by taking the fourth root of the staggered
determinant must be non-local at $a\ne 0$.

It also follows that the staggered theory without
the fourth root
cannot be written as an $SU(4)$-symmetric local theory at $a\ne 0$.
In Ref.~\cite{ah}, it was assumed that Eq.~(\ref{dets}) held with
$\d S_{eff}$ local.\footnote{$\exp(-\d S_{eff})$ was written as $\Det(T)$ in
Ref.~\cite{ah}.}  However, we have shown that such a decomposition is not
possible.

What might be confusing is that the left-hand side of Eq.~(\ref{dets}) is the determinant
of a local operator, $D_{stag}$.  Clearly, the determinant, or equivalently the effective
action $S_{eff}=-\Tr\log(D_{stag})$,  is a non-local object.  What we observe
is simply the fact that the non-locality of $S_{eff}$ cannot be reproduced
entirely by the effective action for the operator $D_{4t}$, because of a conflict between
the symmetries of $D_{stag}$ and $D_{4t}$ at non-zero lattice spacing.
It \textit{is} true that $D_{stag}$ itself can
be written as the sum of taste-invariant and taste-breaking local operators:
\begin{equation}
\label{split}
D_{stag}=D\otimes{\bf 1}+\sum_A D_A\otimes\Xi_A\ ,
\end{equation}
with the $\Xi_A$ a set of fifteen $SU(4)$-algebra valued (hermitian) generators in
taste space,\footnote{We may choose this set to be $\{\x_\m,i\x_\m\x_\n,i\x_\m\x_5,\x_5\}$
with $\x_\m$ a set of $4\times 4$ matrices satisfying $\{\x_\m,\x_\n\}=2\d_{\m\n}$.}
 with $D$ and $D_A$ all local.\footnote{Lattice symmetries,
such as $U(1)_\e$ symmetry, further restrict which $\Xi_A$ can appear, as well
as the form the $D_A$ can take.}
Considering the determinant, however, one has that
\begin{eqnarray}
\label{detcalc}
S_{eff}&=&-\log\Det(D_{stag})\\
&=&-4\log\Det(D)-\log\Det\left(1+\sum_A D^{-1}D_A\otimes\Xi_A\right)\ .\nonumber
\end{eqnarray}
This split of the effective action corresponds to choosing a specific $D$ in Eq.~(\ref{dets}).
Due to
the presence of $D^{-1}$, the second term produces a non-local $\d S_{eff}$, even though the
taste-breaking part of the Dirac operator in Eq.~(\ref{split}) is local.  What we have
argued above, on the basis of the GB spectrum of the staggered theory without
fourth root, is that no split of the form of Eq.~(\ref{dets}) exists for which $\d S_{eff}$
is local.  While it is generally accepted that the taste breaking effects of the
operator $\sum_A D_A\otimes\Xi_A$ vanish in the continuum limit,
it is precisely the non-locality of $\d S_{eff}$ that causes the fifteen GBs
of the staggered theory to be non-degenerate at $a\ne 0$.

While our argument demonstrates that no local lattice theory exists with
a fermion determinant equal to the fourth root of the staggered determinant,
it leaves open the question of whether the non-local behavior
persists in the continuum limit.  Nevertheless, Eq.~(\ref{detcalc}) lends support to the
conjecture that the non-localities vanish in this limit.  Although $\d S_{eff}$ is non-local,
the operator $\sum_A D_A\otimes\Xi_A$ is of
order $a$. Thus
the effects of $\d S_{eff}$ should vanish when
the limit $a\to 0$ is taken while keeping physical momenta
fixed.\footnote{We expect that the continuum limit will have
to be taken before the theory is continued to Minkowski space.}

\vspace{5ex}
\noindent {\large\bf 3.~Examples}
\secteq{3}
\vspace{3ex}

To make the discussion more concrete,
we now give a possible prescription for the construction of the operator
$D_{4t}$ in Eq.~(\ref{split}).
We begin with a massive staggered Dirac
operator
$D_{stag}(m)=D_{stag}(0)+m$  with bare quark mass $m$ in the one-component formalism.\footnote{We will
make the dependence on the quark mass explicit for the rest of this paper.}
There exists a gauge-covariant unitary transformation
$Q^{(0)}$ which puts the theory into the taste representation of Refs.~\cite{gliozzi,saclay}.%
\footnote{The transformation $Q^{(0)}$ is
not unique; see Ref.~\cite{ysrg} for details.}
We may however carry out this transformation as a gaussian renormalization-group (RG)
blocking, leading to a staggered Dirac operator in the taste representation
$D_{taste}(m)$ given by \cite{ys}
\begin{equation}
\label{rgstep0}
D_{taste}^{-1}(m)=\frac{1}{\a}+Q^{(0)}D_{stag}^{-1}(m)Q^{(0)\dagger}\ ,
\end{equation}
where $\a$ is a parameter which appears in the gaussian blocking kernel.
We then have that
\begin{equation}
\label{detsrg}
\Det(D_{stag}(m))=\Det(G^{-1})\;\Det(D_{taste}(m))\ ,
\end{equation}
with
\begin{equation}
\label{G}
G^{-1}=\frac{1}{\a}D_{stag}(m)+Q^{(0)\dagger} Q^{(0)}=\frac{1}{\a}D_{stag}(m)+1\ ,
\end{equation}
where in the last step we have used the fact that the kernel $Q^{(0)}$
is unitary for this ``RG blocking."  For $\alpha\to\infty$, one recovers a transformation of the type
considered in Ref.~\cite{saclay}, but we will take $\a$ to be finite here.
Because $G^{-1}$ is a Dirac operator with a mass of order $\a$ in lattice units,
all the long-distance physics should be contained in
$D_{taste}(m)$.

Again following Ref.~\cite{ys}, one may use $D_{taste}$ as the input for $n$ true RG
blocking steps (in which actual thinning out of fermionic degrees of freedom
occurs) with an RG blocking kernel $Q^{(n)}$. The $n^{\rm th}$ blocking step takes us from a lattice with
spacing $a_{n-1}$ to a lattice with spacing $a_n=2a_{n-1}$;  $a_0$ is defined to be
the spacing of the lattice associated with $D_{taste,0}\equiv D_{taste}$ and is twice
the spacing of the original lattice  on which $D_{stag}$ is defined.
Blocked operators $D_{taste,n}$
and $G_n^{-1}$ result from this process, with, recursively,
\begin{eqnarray}
\label{rgstep}
D_{taste,n}^{-1}(m)&=&\frac{1}{\a}+Q^{(n)}D_{taste,n-1}^{-1}(m)Q^{(n)\dagger}\ ,\\
G_n^{-1}&=&\frac{1}{\a}D_{taste,n-1}(m)+Q^{(n)\dagger}Q^{(n)}\ \nonumber\\
Q^{(n)}Q^{(n)\dagger}&=&c{\bf 1}\ ,\nonumber
\end{eqnarray}
where $c$ is a positive constant, and here ``$\bf 1$" stands for the Kronecker delta
on the coarse lattice.
One expects that the long-distance physics is entirely carried by $D_{taste,n}^{-1}$,
which is manifestly the sum of a smeared quark propagator and a contact term, while
$\Tr\log(G_n^{-1})$ is a local functional of the gauge field.
The determinants are related by
\begin{equation}
\label{detrelation}
\Det(D_{stag}(m))=\Det(D_{taste,n}(m))\prod_{k=0}^n\Det(G_k^{-1})\ ,
\end{equation}
with $G_0^{-1}\equiv G^{-1}$ from Eq.~(\ref{G}).
While Eq.~(\ref{detrelation}) resembles Eq.~(\ref{dets}), it is fundamentally different.  In Eq.~(\ref{detrelation}),
both $\Det(D_{stag}(m))$ and $\Det(D_{taste,n}(m))$ describe the same long-distance
physics, and the factor $\prod_{k=0}^n\Det(G_k^{-1})$ is expected to be a local functional of the
gauge field.  For any finite $n$, both $D_{stag}(m)$ and $D_{taste,n}(m)$ break
taste symmetry, consistent with our general arguments above.

The massless one-component action is invariant under $U(1)_\e$ transformations \cite{ks},
\begin{equation}
\label{u1eps}
\d\c(x)=i\e(x)\c(x)\ ,\ \ \ \ \ \d\overline{\c}(x)=i\e(x)\overline{\c}(x)\ ,
\end{equation}
because $\e(x)\equiv(-1)^{x_1+x_2+x_3+x_4}$ anti-commutes with $D_{stag}(0)$.
From
\begin{equation}
\label{Qtrans}
Q^{(0)}\,\e=(\g_5\otimes\x_5)\,Q^{(0)}\ ,
\end{equation}
it follows \cite{ys} that $D_{taste}=D_{taste,0}$ satisfies a Ginsparg--Wilson (GW) relation \cite{gw}
\begin{equation}
\label{gw}
\{\g_5\otimes\x_5,D_{taste}^{-1}(0)\}=\frac{2}{\a}\;(\g_5\otimes\x_5)\ ,
\end{equation}
if the original operator $D_{stag}$ is massless.\footnote{Note that this reduces to an ordinary
chiral symmetry for $\a\to\infty$.}    Using Eqs.~(\ref{rgstep0}) and \eq{Qtrans} one can show that
$(\g_5\otimes\x_5)D_{taste}(0)$ is hermitian.  Equation~(\ref{gw}) then implies that the
eigenvalues of $D_{taste}(0)$ lie on a circle in the complex plane crossing the real axis
at $0$ and $\a$, with center at $\a/2$.

If we start with a massive staggered Dirac operator
$D_{stag}(m)$ in the one-component formalism,
we obtain a corresponding massive operator $D_{taste}(m)$ in the taste representation.
Using the fact that $D_{taste}(m)=D_{taste}(0)+m$ for $\a=\infty$, it is straightforward
to show for finite $\a$ that
\begin{equation}
\label{Dstm}
D_{taste}(m)=\frac{D_{taste}(0)+m\left(1-\frac{1}{\a}D_{taste}(0)\right)}{1+\frac{m}{\a}
\left(1-\frac{1}{\a}D_{taste}(0)\right)}\ .
\end{equation}
This operator is local, because the second term in the denominator is small
compared to the $1$
(as long as $m\ll 1$ in lattice units).
The eigenvalues still lie on a circle, now with center $(\a/2+m)/(1+m/\a)$ and
radius $(\a/2)/(1+m/\a)$.  In particular, the two possible real eigenvalues are
$m/(1+\frac{m}{\a})$ and $\a$.

In general, $D_{taste,n}(0)$ satisfies a GW relation for any $n$, since the RG kernels $Q^{(n)}$ for $n=1,\dots$
are trivial with respect to Dirac and taste indices.   Explicitly, we have that \cite{ys}
\begin{eqnarray}
\label{gwn}
\{\g_5\otimes\x_5,D_{taste,n}^{-1}(0)\}&=&\frac{2}{\a_n}\;(\g_5\otimes\x_5)\ ,\\
\a_n&=&\frac{1-c}{1-c^{n+1}}\,\a\ .\nonumber
\end{eqnarray}

$D_{taste,n}$ is not invariant under the full taste $SU(4)$ for any finite $n$.
We may construct an $SU(4)$ taste-invariant operator by simply taking the trace in taste space:
\begin{equation}
\label{inv}
D_{inv,n}(m)=\frac{1}{4}\;\tr(D_{taste,n}(m))\otimes{\bf 1}\ ,
\end{equation}
where $\tr$ denotes a trace over taste only.     This operator is not necessarily massless
if we set $m=0$, but whatever quark mass the theory defined by $D_{inv,n}(m)$ has,
it is proportional to the unit matrix in taste space.  It is also clear that
$D_{inv,n}(0)$ does not satisfy a GW relation.

However, it is
straightforward to construct an operator that does obey a GW relation.
In order to do this, we note that $D_{inv,n}$ has no fermion species doublers
for finite $\a$. (We will show this explicitly in Sec.~3a.)
Furthermore, the fact that $\e$ anti-commutes with $D_{stag}(0)$, combined with
anti-hermiticity of $D_{stag}(0)$, implies that
\begin{equation}
\label{g5oc}
\left(D_{stag}^{-1}(m)\right)^\dagger=\e\,D_{stag}^{-1}(m)\,\e\ .
\end{equation}
Using Eqs.~(\ref{rgstep0}), \eq{rgstep}, \eq{Qtrans}  and \eq{inv},
it is then easy to see that $\g_5D_{inv,n}(m)$
is hermitian.  We may thus construct a taste-invariant overlap operator, just as when one
starts with a Wilson--Dirac operator \cite{hn}:
\begin{equation}
\label{ov}
D_{ov,n}\equiv \frac{\a_n}{2}\Bigg(1-\g_5\,{\rm sign}\left(\g_5\Big(1-\frac{2}{\a_n}D_{inv,n}(0)\Big)\right)\Bigg)\ ,
\end{equation}
with $\a_n$ given in Eq.~(\ref{gwn}).
Since this operator is taste invariant, it satisfies a GW relation for any taste matrix $\X$:
\begin{equation}
\label{ovgw}
 \{\g_5\otimes\X,D_{ov,n}^{-1}\}=\frac{2}{\a_n}\;(\g_5\otimes\X)\ .
\end{equation}
It follows that $D_{ov,n}$ is a massless operator.\footnote{This is true even if the original operator $D_{stag}(m)$
is not massless, \ie\ if $D_{inv,n}(0)$ is replaced by $D_{inv,n}(m)$ on the right-hand side of Eq.~(\ref{ov}).}

The operator $D_{ov,n}$ can be written as $D\otimes{\bf 1}$ as in Eq.~(\ref{su4}),
and the resulting $D$ is a possible choice for use in Eqs.~(\ref{4throot}) and \eq{dets}.
Obviously, we can only have that $D_{ov,n}\to D_{taste,n}(m)$ for $n\to\infty$
if we take the original one-component staggered operator to be massless,
so that Eq.~(\ref{gwn}) coincides with Eq.~(\ref{ovgw})
for $\X=\x_5$.
This overlap operator is ``natural,"
because it has been constructed such that the difference between $D_{ov,n}$ and
$D_{taste,n}(0)$ is expected to be of order $a_0^2/a_n^2 = 1/2^{2n}$ \cite{ysrg}.  The distinction is that,
by construction, $D_{ov,n}$ has exact $SU(4)$ taste
symmetry (in fact a full chiral $SU(4)_L\times SU(4)_R$), while $D_{taste,n}(0)$ does not.
The expectation that the difference decreases like
$1/2^{2n}$ arises from the similar expectation that taste symmetry is restored in
the unrooted staggered theory as
we take $n\to\infty$, \ie\ as the lattice spacing of the
original (unblocked) theory is sent to zero.

The sequence of overlap operators can be made massive by choosing
\begin{equation}
\label{ovm}
D_{ov,n}(m) = D_{ov,n}(0) + D_{inv,n}(m) - D_{inv,n}(0)\ ,
\quad\quad m\ne 0\ ,
\end{equation}
with $m$ the original bare staggered mass,
and $D_{ov,n}(0) \equiv D_{ov,n}$ of Eq.~(\ref{ov}).
Our choice is different from the massive overlap operator commonly
used in the literature.  The reason is that, this way, we maintain
the above naturalness property for $m\ne 0$ as well.
For details, see Appendix~A1. Unless  the $n\to\infty$ limit is taken,
the two theories defined by $D_{taste,n}(m)$ and
$D_{ov,n}(m)$ will not have the same renormalized mass; but since the mass in
both theories renormalizes multiplicatively, both theories are massless for $m=0$.
This follows from the fact that both $D_{taste,n}(0)$ and $D_{ov,n}(0)$ have a
Ginsparg--Wilson--L\"uscher (GWL) chiral symmetry \cite{ml}.
Any of the operators $D_{ov,n}(m)$ is a possible choice for $D_{4t}$ in Eq.~(\ref{su4}).

\vspace{5ex}
\noindent {\large\it 3a.~The free case}
\vspace{3ex}

The free case provides an explicit example of
Eq.~(\ref{dets}), with $\d S_{eff}$ non-local.  We choose $n=0$
and use $D_{inv,0}(m)$ for $D_{4t}=D\otimes{\bf 1}$ on the right-hand side of this
equation.  In the free case, a $Q^{(0)}$ exists such that \cite{gliozzi,saclay}
\begin{equation}
\label{sacl}
Q^{(0)}D_{stag}(m)Q^{(0)\dagger}=\sum_\m\bigg(i(\g_\m\otimes {\bf 1})\sin{p_\m}+
(\g_5\otimes\x_\m\x_5)(1-\cos{p_\m})\bigg)+({\bf 1}\otimes{\bf 1}) m
\end{equation}
in momentum space.
Using this $Q^{(0)}$ in Eqs.~(\ref{rgstep0}) and \eq{inv}, we obtain
\begin{eqnarray}
\label{Dstfree}
D_{taste,0}(m)&=&\frac{\sum_\m\bigg(i(\g_\m\otimes {\bf 1})\barp_\m+
\frac{1}{2}(\g_5\otimes\x_\m\x_5)\hatp^2_\m\bigg)+({\bf 1}\otimes{\bf 1})
\left(m+\frac{1}{\a}(\hatp^2+m^2)\right)}{1+\frac{2m}{\a}+\frac{1}{\a^2}(\hatp^2+m^2)}\ ,\nonumber\\
D_{inv,0}(m)&=&\frac{\sum_\m i(\g_\m\otimes {\bf 1})\barp_\m
+({\bf 1}\otimes{\bf 1})\left(m+\frac{1}{\a}(\hatp^2+m^2)\right)}{1+\frac{2m}{\a}+\frac{1}{\a^2}(\hatp^2+m^2)}\ ,
\end{eqnarray}
where
\begin{eqnarray}
\label{ps}
\barp_\m&\equiv &\sin{p_\m}\ ,\nonumber \\
\hatp_\m&\equiv &2\sin{(p_\m/2)}\ , \\
\hatp^2&\equiv &\sum_\m\hatp^2_\m\ .\nonumber
\end{eqnarray}
We see that $D_{inv,0}(m)$ is a Wilson-like Dirac operator, and thus has
no fermion doubling as long as $\a$ is finite.  The massless overlap operator of Eq.~(\ref{ov})
in the free case is
\begin{equation}
\label{ovfree}
D_{ov,0}(0)=\frac{\a}{2}\left(1-\frac{1-\frac{2}{\a}D_{inv,0}(0)}{\sqrt{
1-\frac{\a^2\sum_\m\hatp^4_\m}{(\a^2+\hatp^2)^2}}}\right)
=D_{inv,0}(0)+O(p^4)\ .
\end{equation}
The argument of the square root is strictly positive as long as $\a<2$.

We may now calculate $\d S_{eff}$ for the free case from Eqs.~(\ref{dets}) and \eq{detcalc},
choosing $D\otimes {\bf 1}=D_{inv,0}(m)$ and using Eq.~(\ref{Dstfree}). We find
\begin{equation}
\label{detTfree}
e^{-\d S_{eff}}=
\prod_p\left(1+\frac{\frac{1}{4}\sum_\m\hatp^4_\m}{\barp^2+\left(m+\frac{1}{\a}(\hatp^2+m^2)\right)^2}
\right)^8\ .
\end{equation}
Defining $\d {\cal L}_{eff}$ by $\d S_{eff} = -\Tr(\d {\cal L}_{eff}) $, we have $\d {\cal L}_{eff}(p)
\sim (\sum_\m p^4_\m)/(p^2+m^2)$ at small $p$ (and $am\ll1$). This implies that the Fourier transform
 $\d {\cal L}_{eff}(x-y) $ decays like inverse powers
of the separation $x-y$ (or its components) times a factor $e^{-m|x-y|}$.
Because $m$ is a physical scale, $\d S_{eff}$ is non-local,
Choosing $D=D_{ov,0}(m)$ instead in Eq.~(\ref{dets}) gives a similar
result.  While this only demonstrates the non-locality of $\d S_{eff}$ in the free case,
it is clear  that in the interacting case the non-locality would be gauge-field dependent
(see the discussion around Eq.~(\ref{detcalc})).

If we choose to consider the case of $D_{taste,n}(m)$ for $n>0$, the exact expressions
become more cumbersome.  However, using the free-theory results of \cite{ys}, it
is possible to show that
\begin{equation}
\label{Dstnfree}
D_{taste,n}(m)=\sum_\m\left( i(\g_\m\otimes{\bf 1})p_\m+\frac{1}{2^{n+1}}(\g_5\otimes\x_\m\x_5)p_\m^2+
({\bf 1}\otimes{\bf 1})m+O\bigg(\frac{m^2}{2^n},\frac{p^3}{2^{2n}}\bigg)\right)\ ,
\end{equation}
for small $p$, leading to
\begin{equation}
\label{detTfreen}
\d S_{eff}=-8\sum_p\;\frac{1}{2^{2(n+1)}}\,\frac{\sum_\m p_\m^4}{p^2+m^2}+\dots
\end{equation}
for small $p$.  This shows explicitly how $\d S_{eff}\to 0$ for $n\to\infty$, but also how
$\d S_{eff}$ is non-local for any fixed $n$.

The free case is rather special in that there are no pions, so the
argument of Sec.~2 does not apply. This allows for the possibility
that there may be other choices for the operator $D$ in Eq.~(\ref{dets})
for which $\d S_{eff}$ is local. Indeed, Adams \cite{Adams:2004mf} has constructed
such an operator, which has range $\sqrt{a/m}$ and $\d S_{eff}=0$.  However, the
general features of the GB spectrum show that a similar construction
is not possible in the interacting case.

\vspace{5ex}
\noindent {\large\it 3b.~Background with non-zero topological charge}
\vspace{3ex}

Another example
is provided by the staggered Dirac operator in the
background of a smooth gauge field with fixed topological charge $Q=1$.
Here, we take the operator $D_{4t}=D\otimes{\bf 1}$ of Eq.~(\ref{su4})
to be an overlap operator, and use it (setting $m=0$)
to define the topological charge
of the gauge field under consideration.  As already mentioned,
a possible choice would be one of the $D_{ov,n}$
of Eq.~(\ref{ov}), but in principle any overlap operator will do.

Our choice of gauge field implies that the operator $D$ will have one exact zero mode for quark
mass $m=0$,
and thus one eigenvalue proportional to $m$ when $m\not=0$.
It follows that $\Det^4(D)$ on the right-hand side of Eq.~(\ref{dets}) will be
proportional to $m^4$ and vanish as $m\to 0$.  The operator
$D_{stag}$
on the left-hand side of Eq.~(\ref{dets}) will not have any exact zero modes for a generic gauge-field
configuration (in any topological sector) at non-zero lattice spacing. Instead, it will have four non-degenerate corresponding eigenvalues
\begin{equation}
\label{ev}
\l_i=m+c_i a^\g\ ,\ \ \ \ \ i=1,\dots,4\ ,
\end{equation}
with $\g$ a positive exponent.\footnote{For generic eigenvalues,
one expects that $\g=1$.  It could be that $\g=2$ for zero modes.  The precise value
of $\g$ does not affect our argument.}
In general none of the $c_i$ will be exactly zero:
If we consider for instance an instanton
with radius $\r$, the $c_i$ will be proportional to $\r^{-\g-1}$.  It follows that
\begin{equation}
\label{detTinst}
e^{-\d S_{eff}}\propto\prod_i\left(1+\frac{c_i a^\g}{m}\right)\ .
\end{equation}
This is just the zero-mode contribution to
\begin{equation}
\label{Seffinst}
\d S_{eff}=-\Tr\log\left(1+ \sum_A D^{-1}D_A\otimes\X_A\right)\ ,
\end{equation}
in Eq.~(\ref{detcalc}), where now $D$ has been chosen to be an overlap operator.\footnote{
In this case the sum over $A$ includes a term with $\X_A={\bf 1}$.}
Thus, the $1/m$ signals the dependence of $\d S_{eff}$ on the non-local $D^{-1}$.
Note also that $\d S_{eff}$ diverges in the chiral limit for any non-zero
lattice spacing,   exhibiting the well-known fact that the
chiral and continuum limits do not commute \cite{cbcl}.

\vspace{5ex}
\noindent {\large\it 3c.~Consequences for Reference~\cite{ah}}
\vspace{3ex}

Our results invalidate the basic assumption made in Ref.~\cite{ah},
which was that $\delta S_{eff}$ defined by Eq.~(\ref{dets}) cannot affect long-distance physics
even at non-zero $a$.  Instead, we find that $\delta S_{eff}$ has to contain long-distance
physics at $a\ne 0$ because of the mismatched symmetries of $D_{stag}$
and $D_{4t}$ of Eqs.~(\ref{dets}) and \eq{su4}.   Contrary to what was suggested
in Ref.~\cite{ah}, it is not possible to reconcile the theories described by
$D_{stag}$ and $D_{4t}$ by an additive shift in the quark mass.
Unlike the theory defined by $D_{stag}$,
the theory defined by $D_{4t}$ has to contain fifteen
(pseudo) Goldstone bosons, which remain degenerate even if they pick up a mass due to the presence of
an explicit ($SU(4)$-symmetric) quark mass.
Based on a comparison of zero modes of $D_{stag}$ and $D_{4t}$,
Ref.~\cite{ah} furthermore argues that
the quark masses of the two theories have to be related
by an $O(a^2)$ additive quark mass renormalization. That argument fails, however,
precisely because the relation between the two theories is non-local.
Our construction of the overlap operators $D_{ov,n}(m)$
demonstrates that in fact $SU(4)$-symmetric lattice Dirac operators exists which
becomes exactly massless when the staggered quark mass $m$ is set equal to zero.
We emphasize however that any overlap operator can be used to invalidate the
claim of Ref.~\cite{ah}, as discussed in Sec.~3b.
A correct description of the approach of the continuum limit as far as the
physics of GBs is concerned is provided by staggered chiral perturbation theory
\cite{ls,ab}.

\vspace{5ex}
\noindent {\large\bf 4.~Conclusion}
\secteq{4}
\vspace{3ex}

Our main result is a proof in Sec.~2 that the theory defined by the fourth root of the
staggered fermion determinant does not correspond to a local theory at non-zero
lattice spacing $a$.  This follows from the fact that $SU(4)$ taste symmetry is broken
at non-zero $a$ in the unrooted staggered theory.  If a local theory corresponding
to the fourth-root theory existed, one could take four copies
of it and construct  a local
theory with exact $SU(4)$ taste symmetry, {\it cf.}\ the theory defined by $D_{4t}$ in Eq.~(\ref{su4}).
The $SU(4)$ symmetry implies that the fifteen pseudo-Goldstone bosons of this
theory must be degenerate.  On the other hand,
it is well known that
the 15 pseudo-Goldstone bosons in the staggered theory at non-zero $a$ are non-degenerate
because of taste violations.
There is thus a mismatch
in the long-distance physics
of the staggered and $SU(4)$ theories when $a\not=0$.
The contradiction implies that the rooted theory cannot be local
at non-zero $a$:  $\d S_{eff}$, defined through
Eq.~(\ref{dets}), must be non-local.

The key issue is then whether the non-locality
persists in the continuum limit.  While this remains an open question, the argument
given around Eq.~(\ref{detcalc}) suggests that the theory is in the desired universality class
as long at the continuum limit is taken before the chiral limit. In other words,
it appears that locality will be restored for $a\to0$ at any $m\ne 0$ ({\it cf.} Sec.~3b).
For recent theoretical results supporting this
conjecture, we refer to Refs.~\cite{ysrg,cb}.

While the main argument summarized above stands alone, we have discussed two examples
that make our reasoning more concrete. The examples are provided by the
staggered theory in the free case (Sec.~3a) and in the background of a smooth
gauge field with non-zero topological charge (Sec.~3b).  Starting from
the staggered Dirac operator, we constructed a sequence of overlap operators
$D_{ov,n}$ in Sec.~3, which can be used to give a fermionic definition of
topological charge suited to our arguments.  In both examples, we find that
$\d S_{eff}$ is explicitly non-local, but that the non-local behavior disappears
in the continuum limit.

\vspace{3ex}
\noindent {\bf Note Added}
\vspace{3ex}

Recently, Hasenfratz and Hoffmann \cite{Hasenfratz:2006nw} have posted a paper
that discusses staggered fermions in the context of the
Schwinger model.  They present numerical evidence that the
staggered determinant (on both unrooted and rooted ensembles) can be made approximately equal
to an overlap determinant
by adjusting the overlap mass appropriately, up to a local effective action. When the quark
mass is large
compared to the taste violations, it is not inconsistent
with the arguments given here that the physics
of the overlap and staggered fermions could be approximately
the same.  However, at low quark mass the properties of the GBs
guarantee that the physics of the two theories must be drastically different;
indeed, numerically the matching of determinants deteriorates.
In QCD, current simulations \cite{numev} are
in this ``low mass'' region ($m\sim a^2\Lambda_{QCD}^3$), where staggered
chiral perturbation theory \cite{ls,ab,cb} is the appropriate tool.

\vspace{3ex}
\noindent {\bf Acknowledgments}
\vspace{3ex}

We thank Anna Hasenfratz and Tom  DeGrand for discussions, and are grateful
to the Institute for Nuclear Theory at the University of Washington
for its hospitality.
MG also thanks the Physics Department of the University of Rome ``La Sapienza"
for hospitality.
MG was supported in part by the Generalitat de Catalunya under the program
PIV1-2005; both CB and MG were supported in part  by the US Department of Energy.
YS was supported by the Israel Science Foundation under grant no.~173/05.

\newpage
\vspace{5ex}
\noindent {\large\bf Appendix A.~Selected properties of taste-basis Dirac
operators}
\secteq{A}
\vspace{3ex}

In this appendix we collect a number of useful results pertaining
to the three families of taste-basis Dirac operators considered in the
text: $D_{taste,n}(m)$, $D_{inv,n}(m)$, and $D_{ov,n}(m)$.

\vspace{5ex}
\noindent {\large\it A1.~Construction of $D_{ov,n}(m)$}
\vspace{3ex}

Consider a massless overlap operator $D_{ov}$ that satisfies the GW relation
\begin{equation}
\label{GWagain}
  \{ \g_5, D_{ov} \} = {2\over \a} D_{ov} \g_5 D_{ov}\,.
\end{equation}
Here $\a=O(1/a)$, where $a$ is the lattice spacing.
The choice of a massive overlap operator most common
in the literature is
\begin{equation}
\label{Dovm}
  D_{ov}(m) = (1-m/\a)D_{ov} + m\,,
\end{equation}
where $D_{ov}(0)=D_{ov}$ is a solution of Eq.~(\ref{GWagain}).
In fact, as we will explore,
there is a large freedom in extending the definition of an
overlap operator to the massive case.

Let us spell out the requirements that
a massive overlap operator should meet.  First, the definition (\ref{Dovm})
satisfies
\begin{equation}
\label{sysDov}
  D_{ov}(m) = D_{ov}(0) + Z m + O(m^2a,mpa)\,.
\end{equation}
This is an obvious requirement for any sensible $D_{ov}(m)$.
The $O(m^2a,mpa)$ irrelevant terms
cannot re-introduce any fermion doublers because $ma \ll 1$. Since $m$
is a bare mass, we have allowed for an $O(1)$ multiplicative
renormalization factor $Z$.  In the case of Eq.~(\ref{Dovm}) one has $Z=1$,
but, anticipating less explicit definitions, there is nothing
wrong in principle with having $Z \ne 1$.  Either way, the value of $m$
must be adjusted to reproduce the desired renormalized mass.

The second requirement
has to do with the algebraic transformation properties
under the GWL chiral symmetry \cite{gw,ml}
(for reviews see Refs.~\cite{fn98,mg2000}).  The GW relation (\ref{GWagain})
implies that the operator
\begin{equation}
\label{ghat5}
  \ghat = \g_5 (1 - (2/\a) D_{ov})\,,
\end{equation}
satisfies $\ghat^2=1$. In all relevant cases it will further be true
that $\ghat$ (or its generalization) is hermitian.
A possible choice of the GWL
chiral transformation is then given by $\d\psi = \ghat \psi$,
$\d\psibar = \psibar \g_5$.  The GW relation can be rewritten as
$\g_5 D_{ov} + D_{ov} \ghat =0$, which implies that the fermion action $S_{ov} = \psibar D_{ov} \psi$
is invariant under the GWL transformation (see also Sec.~A3 below).

In the massive case the fermion action cannot be invariant
under the GWL transformation. Instead, in analogy with an ordinary mass term,
and assuming that parity is a symmetry, one requires that the mass term
be a scalar density that transforms into a pseudo-scalar density
under the GWL transformation.
In fact, this requirement can be rather trivially satisfied.
Consider a general bilinear fermion action $S_F = \psibar D \psi$,
assuming only that $S_F$ is hypercubic and parity invariant. Assume also
a given GW operator $D_{ov}$ (with in general $D_{ov} \ne D$).
We introduce the standard chiral projectors $P_{R,L} = \half(1\pm\g_5)$
as well as ``hatted'' chiral projectors
$\Phat_{R,L} = \half(1\pm\ghat)$, and define
$\psi_{R,L} = \Phat_{R,L} \psi$,  $\psibar_{R,L} = \psibar P_{L,R}$.
Note that hatted projectors are used for $\psi$ while ordinary projectors
are used for $\psibar$.
One can now split the action into two parts,
\begin{equation}
\label{Schim}
  S_F = \psibar (D_\c + D_{mass}) \psi \,,
\end{equation}
where
\begin{eqnarray}
\label{Dchi}
  D_\c &=& P_R D \Phat_L + P_L D \Phat_R \,,
\\
  D_{mass}  &=& P_R D \Phat_R + P_L D \Phat_L \,.
\label{Dmass}
\end{eqnarray}
Under the chiral GWL transformation, $D_\c$ is invariant,
whereas $D_{mass}$ transforms as required for a mass term.

While the decomposition (\ref{Schim}) is possible for {\it any} $D$,
clearly this does not imply that any $D$ would qualify as a massive
overlap operator.  In accordance with Eqs.~(\ref{sysDov}) and (\ref{Schim}),
we require that a massive overlap operator satisfy
\begin{eqnarray}
\label{Dovchi}
  D_{ov,\c}(m) &=& D_{ov}(0) + O(mpa) \,,
\\
  D_{ov,mass}(m)  &=& Z m + O(m^2a,mpa)\,,
\label{Dovmass}
\end{eqnarray}
where $D_{ov,\c}(m)$ and $D_{ov,mass}(m)$ are defined by substituting
$D_{ov}(m)$ into Eqs.~(\ref{Dchi}) and (\ref{Dmass}) respectively.\footnote{
  The hatted projectors are always defined with respect to $D_{ov}(0)=D_{ov}$.
In the case of Eq.~(\ref{Dovm}) one has $D_{ov,\c}(m)=D_{ov}$
and $D_{ov,mass}(m) = m (P_R \Phat_R + P_L \Phat_L) = m + O(mpa)$.
}
Note that corrections of $O(m^2a)$ are absent in Eq.~(\ref{Dovchi}) because
the difference between $\hat \gamma_5$ and $\gamma_5$ is $O(pa)$, and ordinary
chiral symmetry (as opposed to the GWL type) would forbid mass terms in $D_{ov,\c}(m)$.
Like Eq.~(\ref{sysDov}), this asserts that $D_{ov}(m)$ satisfies a GW relation
in the limit $m\to 0$;  that the difference $D_{ov}(m)-D_{ov}(0)$ is $O(m)$;
and that to leading order, this difference is actually linear in $m$.
What Eqs.~(\ref{Dovchi}) and (\ref{Dovmass}) add is that $D_{ov,mass}(m)$
transforms as expected under the GWL symmetry; the above discussion
clarifies that this additional requirement can always be met for
any operator that already satisfies Eq.~(\ref{sysDov}).
These properties ensure that the mass parameter will be renormalized
multiplicatively.\footnote{
  The (finite) ratio of continuum and lattice $Z$ factors
(both evaluated at the same scale)
will generically be a function of $ma$.
}

Here, we will add one new requirement.
Under a certain scaling assumption to be discussed in Sec.~A2, we demand that
the sequences $D_{inv,n}(m)$ and $D_{ov,n}(m)$ both
have the same $n\to\infty$ limit as the original RG-blocked operators
$D_{taste,n}(m)$, for any $m$.  With $D_{ov,n}(0)=D_{ov,n}$ of Eq.~(\ref{ov}),
a massive overlap operator that satisfies all the above requirements is
(the following is identical to Eq.~(\ref{ovm}) in the main text)
\begin{equation}
\label{Dovmn}
  D_{ov,n}(m) = D_{ov,n}(0) + D_{inv,n}(m) - D_{inv,n}(0)\,,
  \quad\quad  m \ne 0\,.
\end{equation}
Of course, we now define the GWL transformation and the hatted projectors
using $\hat\g_{5,n} = \g_5 (1 - (2/\a_n) D_{ov,n}(0))$.
Equations (\ref{Dovchi}) and (\ref{Dovmass}) follow because,
similarly to Eq.~(\ref{sysDov}),
one has $D_{inv,n}(m)=D_{inv,n}(0) + Z m + O(m^2a,mpa)$.
Note that the proportionality constant $Z$ is necessary in this case, because
$D_{inv,n}(m)$ was defined such that $m$ is the value of the mass
in the original one-component staggered operator.
The $n\to\infty$ convergence properties
will be established in the following subsection.

Last, we briefly comment on the construction
of the low-energy effective theories:
the Symanzik action and the chiral lagrangian.  In the case of Eq.~(\ref{Dovm}),
the GW chiral lagrangian has the same internal symmetries as the continuum
chiral lagrangian.  The situation is slightly more involved in the
more general case of Eqs.~(\ref{Dovchi}) and (\ref{Dovmass}). There, terms proportional
to powers of $ma$ appear in the chirally invariant part of the Dirac operator,
$D_{ov,\c}(m)$. This feature will carry over to the chirally invariant
part of the Symanzik action. In constructing the corresponding chiral
theory, one therefore has to include a chirally invariant spurion
proportional to $ma$.  The spurion would, in effect, make
the low-energy constants (LECs) of the chiral
theory functions of $am$.
 Such mass dependence in the LECs could present a practical
difficulty in extracting chiral physics from a simulation that used $D_{ov}(m)$ as the
fundamental Dirac operator. However, there is no theoretical problem in considering
$D_{ov}(m)$, and
all the standard implications of chiral symmetry are preserved.
In particular, the masses of Goldstone pions vanish in the chiral limit,
for any value of the lattice spacing.

\vspace{5ex}
\noindent {\large\it A2.~Scaling and convergence for $n\to\infty$}
\vspace{3ex}

A basic hypothesis of the RG treatment of
staggered fermions (with or without the fourth root)
is that the taste-breaking terms of the RG-blocked
operator $D_{taste,n}(m)$
tend to zero in the limit of infinitely many RG blocking steps \cite{ys,ysrg}.
The taste-breaking part $\D_n$ is given explicitly by writing
\begin{equation}
\label{Delta}
  D_{taste,n}(m) = D_{inv,n}(m) + \D_n(m) \,,
\end{equation}
where $D_{inv,n}(m)$ is given by Eq.~(\ref{inv}).
We will hold fixed the coarse-lattice spacing $a_c \equiv a_n$
obtained after $n$ blocking steps, implying that $a_0 = 2^{-n} a_n$ goes to zero
when $n$ is taken to infinity.
In the free theory \cite{ys}, one can prove that $\| a_c \D_n \| = O(2^{-n})$.
In the interacting case no proofs can be given;
{\it we will assume that $\D_n$ scales in the same way, up
to logarithmic corrections in} $a_0/a_c$ (that we suppress below).
We refer to Ref.~\cite{ysrg}
for a discussion of the status of this assumption, as well as a more
precise statement about the gauge fields for which it is expected
to apply.

Under this scaling hypothesis it is trivial that
$D_{taste,n}(m)$ and $D_{inv,n}(m)$ have a common $n\to\infty$ limit,
for any $m$. Furthermore, by Eq.~(\ref{Dovmn}), the same will be true for
$D_{ov,n}(m)$, provided $D_{ov,n}(0)$ has the same $n\to\infty$ limit as
$D_{inv,n}(0)$.  We will now prove this.
In the rest of this subsection we set $m=0$ and drop the mass argument.
We begin by substituting Eq.~(\ref{Delta}) into
\begin{eqnarray}
\label{gwn-inv}
\{\g_5\otimes\x_5,D_{taste,n}\}&=&\frac{2}{\a_n}\;D_{taste,n}\,(\g_5\otimes\x_5)\,D_{taste,n}\ ,
\end{eqnarray}
which is equivalent to Eq.~(\ref{gwn}).
We then multiply both sides of the resulting equation by ${\bf 1} \otimes \x_5$,
take the trace over taste indices only, and
form the tensor product with an arbitrary taste matrix $\X$,
obtaining
\begin{eqnarray}
\label{splitGW}
  \{ D_{inv,n}\,, (\g_5 \otimes \X) \}
  - \mbox{} && \hspace{-5.5ex}
  {2\over\a_n} D_{inv,n} (\g_5 \otimes \X) D_{inv,n} \; =
\nonumber\\
  &=& \!
  {1\over 2\a_n}\; \tr\left(
  ({\bf 1} \otimes \x_5) \D_n (\g_5 \otimes \x_5) \D_n \right)
  \otimes \X \,.
\end{eqnarray}
We used that $\D_n$ is traceless on the taste index (compare Eq.~(\ref{inv})).
By the scaling hypothesis,
the right-hand side of Eq.~(\ref{splitGW}) is $O(2^{-2n})$,
which tells us by how much $D_{inv,n}$ fails to satisfy
the GW relation (\ref{ovgw}). Now introducing
\begin{equation}
\label{gtilde}
  \tilde\g_{5,n} = \g_5  (1 - (2/\a_n) D_{inv,n})\,,
\end{equation}
it follows from Eq.~(\ref{splitGW}) that $\tilde\g_{5,n}^2=1+O(2^{-2n})$.
Hence, $\hat\g_{5,n} \equiv {\rm sign}(\tilde\g_{5,n})
= \tilde\g_{5,n}+O(2^{-2n})$.
Finally, inserting this into Eq.~(\ref{ov})
we find $D_{ov,n}=(\alpha_n/2)(1-\g_5\ghat)=D_{inv,n}+O(2^{-2n})$.

\vspace{5ex}
\noindent {\large\it A3.~Index of $D_{taste,n}$}
\vspace{3ex}

Here we address the following issue.  The one-component staggered
theory has an exact chiral symmetry for $m=0$, the $U(1)_\e$ symmetry.
The corresponding chiral transformations
of the continuum four-taste theory are generated by $\g_5 \otimes \x_5$, and they form a non-anomalous subgroup
of $SU(4)_L \times SU(4)_R$.\footnote{
  This is true when the staggered mass term is introduced
  as $D_{stag}(m)=D_{stag}(0)+m$ \cite{mgjs}.
}
In contrast, after any number of RG blocking steps,
we obtain the operator $D_{taste,n}(m)$ which, for $m=0$, only satisfies
the GW relation (\ref{gwn}).
While the RG-blocked action is invariant under the corresponding GWL
transformation, this is not enough to establish that it is a symmetry.
One must further check that the measure term, arising from this change
of variables, vanishes. Here we show that this is indeed the case.
Again we will set $m=0$ and drop the mass argument.

The variation of the measure is given by \cite{ml}
\begin{equation}
\label{msr}
  -\tr((\g_5 \otimes \x_5) D_{taste,n}/\a_n) = {\rm index}(D_{taste,n})\,.
\end{equation}
We note that, loosely speaking, one expects the index of $D_{taste,n}$
to vanish in the continuum limit, because taste symmetry
is recovered in this limit, and $\tr (\x_5)=0$.
We will establish the stronger result
that the index of $D_{taste,n}$ is actually zero on the lattice.
The precise statement is that the index is zero except possibly
on a subspace $\cu_{00} \subset \cu_0 \subset \cu$,
where $\cu$ is the (finite volume) gauge-field space,
$\cu_0$ is the (proper) subspace where $D_{taste,n}$ has at least one exact
zero mode, and $\cu_{00}$ is a proper subspace of $\cu_0$ defined below.
Further, $\cu_0$ is a measure zero subset of $\cu$, and
$\cu_{00}$ is a measure zero subset of $\cu_0$.

One can always choose a basis for the exact zero modes of $D_{taste,n}$ such
that each zero mode $\psi_0$ has a definite chirality,
\begin{equation}
\label{psi0}
  \widehat{(\g_5 \otimes \x_5)} \psi_0 = (\g_5 \otimes \x_5) \psi_0
  = \pm \psi_0 \,,
\end{equation}
where, analogous to Eq.~(\ref{ghat5}),
\begin{equation}
\label{ghatgen}
  \widehat{(\g_5 \otimes \x_5)}
  = (\g_5 \otimes \x_5)  (1 - (2/\a_n) D_{taste,n}) \,.
\end{equation}
By Eq.~(\ref{ghatgen}), ordinary and hatted projectors
coincide when acting on a zero mode.
Also, on a zero mode, the Dirac operator $D_{taste,n}$
commutes with the chiral generator, as usual.  Therefore it is enough
to show that the index of $D_{taste,n}$ is zero with respect
to $\g_5 \otimes \x_5$ chirality.  This can be done by relating the zero modes
of $D_{taste,n}$ to those of $D_{stag}$.  Iterating Eq.~(\ref{rgstep}) we have
\begin{equation}
\label{ttos}
  D_{taste,n}^{-1} = 1/\a_n + Q_n D_{stag}^{-1} Q_n^\dagger \,,
\end{equation}
where $Q_n = Q^{(n)}  Q^{(n-1)} \cdots Q^{(1)}  Q^{(0)} $.
If we gradually vary the gauge field so as to approach a configuration
where $D_{taste,n}$ has an exact zero mode, the norm
of $D_{taste,n}^{-1}$ on the left-hand side diverges.
This is possible only if the norm of $D_{stag}^{-1}$ diverges too.
Thus, not surprisingly, any exact zero mode of $D_{taste,n}$ must be obtained
via RG blocking from an exact zero mode of $D_{stag}$.

Because of $U(1)_\e$ symmetry, the spectrum of $D_{stag}$ consists
of imaginary pairs $\pm i\l$, and the corresponding eigenmodes
are related by multiplication with $\e(x)$. Since the eigenvalues are continuous
functions of the gauge fields, and since there are no zero modes in the
free case, any zero modes that appear must also
be paired.  We choose a chiral basis for the two zero modes, which is always possible. Then,
as the gauge field changes, the off-diagonal matrix element of $D_{stag}$ between
the modes is not forbidden by  $U(1)_\e$ symmetry,
and is thus generically non-zero.  This suggests  --- in accordance with standard lore ---
that exact zero modes exist only on a zero measure subspace $\cu_0$.

Using Eq.~(\ref{Qtrans}) it follows from
the above discussion that, given a pair of zero modes of $D_{stag}$,
then $D_{taste,0}$ must have a corresponding pair of zero modes,
with one zero mode of each $\g_5 \otimes \x_5$ chirality.
The index of both $D_{stag}$ and $D_{taste,0}$ is, thus, {\it always} zero.
The index of  $D_{taste,n}$ could only be non-zero
if the blocking transformation $ Q^{(n)}  Q^{(n-1)} \cdots Q^{(1)}$
exactly annihilated one of the definite-chirality zero modes
of $D_{taste,0}$ but not the other.  Generically this will
not happen, and the subspace $\cu_{00}$ where this does happen
therefore has measure zero with respect to $\cu_0$.
(We leave it open whether or not $\cu_{00}$ is an empty set.)
Assuming that no (interesting) QCD observable has a $\d$-function support on $\cu_{00}$,
the GWL transformation is then a symmetry of the RG blocked theory.


\end{document}